# Decoding the circuitry of consciousness:

# from local microcircuits to brain-scale networks

Julien Modolo[1], Mahmoud Hassan[1], Fabrice Wendling[1,*], and Pascal Benquet[1]

1: Univ Rennes, INSERM, LTSI – U1099, F-35000 Rennes, France

*Corresponding author: Fabrice Wendling – fabrice.wendling@univ-rennes1.fr

## Abstract

Identifying the physiological processes underlying the emergence and maintenance of consciousness is one of the most fundamental problems of neuroscience, with implications ranging from fundamental neuroscience to the treatment of patients with disorders of consciousness (DOC). One major challenge is to understand how cortical circuits at drastically different spatial scales, from local networks to brain-scale networks, operate in concert to enable consciousness, and how those processes are impaired in DOC patients. In this review, we attempt to relate available neurophysiological and clinical data with existing theoretical models of consciousness, while linking the micro- and macro-circuit levels. First, we address the relationships between awareness and wakefulness on the one hand, and cortico-cortical, and thalamo-cortical connectivity on the other hand. Second, we discuss the role of three main types of GABAergic interneurons in specific circuits responsible for the dynamical re-organization of functional networks. Third, we explore advances in the functional role of nested oscillations for neural synchronization and communication, emphasizing the importance of the balance between local (high-frequency) and distant (low-frequency) activity for efficient information processing.

The clinical implications of these theoretical considerations are presented. We propose that such cellular-scale mechanisms could extend current theories of consciousness.

**Keywords**

Disorders of consciousness, functional connectivity, micro-circuitry, communication through coherence, gating by inhibition, electroencephalography.

**Introduction**

Understanding how consciousness arises from communication among brain regions is a question of the utmost importance in the field of neuroscience in general, and for the diagnosis and treatment of patients suffering from disorders of consciousness (DOC) in particular. The problem of consciousness can be seen as fundamental (e.g.: "What is consciousness? Why do we have subjective, conscious experiences?", such questions are referred to as the "hard" problem of consciousness (Harnad, 1998)) or more empirical (e.g.: "What are the processes associated with the emergence and maintenance of consciousness?", this forms the "soft" problem of consciousness (Harnad, 1998)). In this review, we aim at understanding 1) how brain networks at different scales are involved in enabling and maintaining conscious processes of information transmission and processing related to awareness and wakefulness, and 2) how these mechanisms are related to the disruptions of consciousness in DOC patients.

Many theories have been proposed to explain how consciousness originates, ranging from abstract and informational concepts to neurophysiology-based theories. The most widespread theories of consciousness have a fundamental assumption in common: information processing in the human brain networks is inextricably linked with consciousness. A recent paper by Dehaene and colleagues summarizes this principle as follows (Dehaene et al., 2017):

*"What we call "consciousness" results from specific types of information-processing computations, physically realized by the hardware of the brain."*

The three main theories of consciousness include the Integrated Information Theory (IIT) (Tononi, 2004), the Dynamic Core Hypothesis (Tononi and Edelman, 1998) (DCH), and the Global Workspace Theory (Baars, 1988; Dehaene et al., 1998; Dehaene et al., 2003) (GWT). Historically, DCH theory has been the first to refer to the notion of information processing involved in consciousness (Tononi and Edelman, 1998). This theory is based on the central role of functional clusters in the thalamo-cortical system and re-entrant interactions, with high integration and differentiation of neuronal activity being crucial in the emergence of conscious phenomena. IIT, which is an evolution and generalization of DCH, is based on a set of axioms from which postulates are derived. IIT also provides a computable quantity, $\Phi$, also called *integrated information*, that quantifies the level of consciousness. In this framework, if combining sub-elements increases information processing capability more than linearly adding these elements, then integrated information increases. Global Workspace Theory (GWT) is a theory of consciousness theory that is more directly connected with neurophysiology and neuroanatomy. The main hypothesis of GWT is that conscious information is globally available within the brain, and that two fundamentally different computational systems co-exist: 1) a network of distributed "local" processors operating in parallel in the brain ("unconscious"), and 2) a "global" workspace formed by a network of distributed interconnected cortical areas involved in conscious perception (Baars, 1988). The key concept here is that the global workspace is composed of distant regions densely connected through glutamatergic cortico-cortical connections as opposed to the network of local processors operating in "isolation" (in parallel). It is worth noting that this distinction between unconscious and conscious processes has been recently challenged, and might be an oversimplification (Melnikoff and Bargh, 2018). In GWT, conscious perception is associated with "ignition", a large-scale brain activation pattern induced by exposure to a stimulus (Dehaene et al., 2003). If the stimulus does not trigger ignition, and if the induced brain response remains spatially confined and is brief, then the perception will not reach consciousness. In other terms, a stimulus has to be sufficiently long and strong to reach consciousness, which suggests a form of filtering mechanism that is consistent with the view that only a limited amount of information effectively enters in the global workspace. Despite these successes in accounting for experimental data in humans regarding subliminal (unconscious) and conscious perception (Sergent and Dehaene, 2004; King et al., 2016), one drawback of GWT is that it does not explicitly relate the large-scale recruitment of brain regions during conscious access with cellular mechanisms. More precisely, what prevents ignition for short, irrelevant stimuli; and conversely, what enables ignition for strong stimuli? The neuroanatomical, neurophysiological and dynamical mechanisms behind ignition are of fundamental importance to understand how we become conscious of a stimulus, or how alterations of brain networks can result in impaired consciousness in DOC patients.

If one accepts that consciousness is associated with a sufficiently complex (in the algorithmic sense of "less compressible") information processing, then the emergence of consciousness is critically dependent on three factors: 1) a physical network enabling interactions between its components; 2) the flexibility to re-organize transiently sub-networks to achieve greater computation capabilities by increasing the number of possible configurations and input-output functions, through functional connectivity; and 3) dynamic communications between its components. These three critical components have the potential to be altered, for example in lesions following traumatic brain injury. While the physical large-scale network linking brain regions is well defined and known as the connectome, there are still unresolved questions regarding the transient organization of clusters performing specific computations (functional networks), the associated means of communication (neural coding) and how large-scale functional brain networks and information routing can reconfigure rapidly depending on micro-circuits regulation.

In this review, the objective is therefore to propose a bridging between cortical microcircuits on the one hand (cellular scale), and brain-scale activity associated with the two main dimensions of conscious perception (awareness and wakefulness) on the other hand. Such multi-scale understanding is a pre-requisite to understand how brain networks become dysfunctional in DOC, and might contribute to reconcile GWT and DCH into a unified framework.

The review is organized as follows. First, we examine the relationship between the two dimensions of consciousness, namely wakefulness and awareness, with functional connectivity between cortical regions and the thalamus. Second, we review the "means of communication" enabling complex information processing linked with consciousness, which regulate cortico-cortical communication, among which communication through coherence (CTC) and gating by inhibition (GBI). The alteration of those mechanisms is presented through results from the clinical literature. Third, we attempt at linking these findings with concepts that have recently emerged based on the communication between brain regions based on cross-frequency couplings between oscillations with specific functional roles. Finally, we suggest possible clinical implications of this framework in terms of novel neuromodulation protocols in DOC.

## 1. Awareness, wakefulness: a short review of concepts

Conscious perception results from an interplay between two processes interacting with each other: *awareness* and *wakefulness*. Deep sleep switches awareness off, whereas being able of conscious perception (awareness) of environmental stimuli usually implies a state of wakefulness, as illustrated in Figure 1. For example, during general anaesthesia, there is both an absence of conscious perception, awareness and wakefulness. In some peculiar cases, however, these two components can be unrelated. Unresponsive Wakefulness State (UWS) is an example of Disorder of Consciousness (DOC) in which wakefulness is present without any detectable signs of awareness (Laureys and Boly, 2012). Also, during lucid dreaming, there is a form of awareness in the absence of wakefulness (during sleep) (Voss et al., 2013). Another example is spatial neglect syndrome, in which patients have no conscious awareness of visual stimuli, while being awake in the contralateral side of the cortical lesion (Le et al., 2015).

*Awareness* is supported by attentional, fronto-parietal networks that amplify synaptic connections within specific cortical pathways (Tallon-Baudry, 2011). This amplification of relevant stimuli enhances the activated network related to stimulus representation. In parallel, the concomitant inhibition of irrelevant surrounding networks i) optimizes cortico-cortical routing of information by constraining the possible propagation of neural activity throughout all possible cortical "routes", which ii) restricts propagation to a limited number of stimulus-driven possibilities, and iii) increases the signal-to-noise ratio. Such mechanisms are related to the concept of functional connectivity, and are detailed further in this review. Importantly, such mechanisms of active inhibition likely involve cortical inhibition with an active modulation by thalamocortical inputs (Gabernet et al., 2005), implying that the pattern of thalamo-cortical activity influences information processing in cortico-cortical networks.

*Wakefulness* depends critically on thalamo-cortical connectivity and neuromodulatory brainstem inputs to the thalamus (including noradrenaline projections from the locus coeruleus (Monti, 2011)). For instance, during slow-wave sleep, a low-frequency, synchronized activity between the cortex and thalamus (the so-called "up-and-down" rhythm (Neske, 2015), prevents transmission of sub-cortical inputs to the cortex during sleep). This provides an example in which thalamo-cortical drastically decreases information processing by cortico-cortical networks, resulting in a loss of consciousness. During wakefulness, thalamo-cortical activity is weakly synchronized (Gent et al., 2018), which is a necessary, but not sufficient condition to enable consciousness. For example, as aforementioned, wakefulness is present in UWS patients but cortico-cortical communication is severely impaired (Noirhomme et al., 2010), interfering with the "awareness" component of consciousness. Another required condition for consciousness is an efficient large-scale cortico-cortical communication that can support awareness through the activation of attentional fronto-parietal networks (Luckmann et al., 2014; Ptak et al., 2017). Therefore, in terms of neuroanatomy, it is possible to link wakefulness with thalamo-cortical, "vertical" connectivity, while awareness depends on cortico-cortical, "horizontal" connectivity. This is consistent with the recent view by Naccache (Naccache, 2018) that Minimally Conscious State (MCS) patients, who are conscious to some degree, exhibit "Cortically Mediated States", whereas UWS patients do not exhibit such activity, presumably because cortico-cortical connectivity (critical for awareness) is too severely impaired. More specifically, a "critical mass" of information processing requires occurring for the emergence and maintenance of consciousness, which is tightly regulated by thalamo-cortical and cortico-cortical connectivity. Figure 1 presents, in a two-dimensional plane, the continuum of the states of consciousness, as a function of awareness and wakefulness.

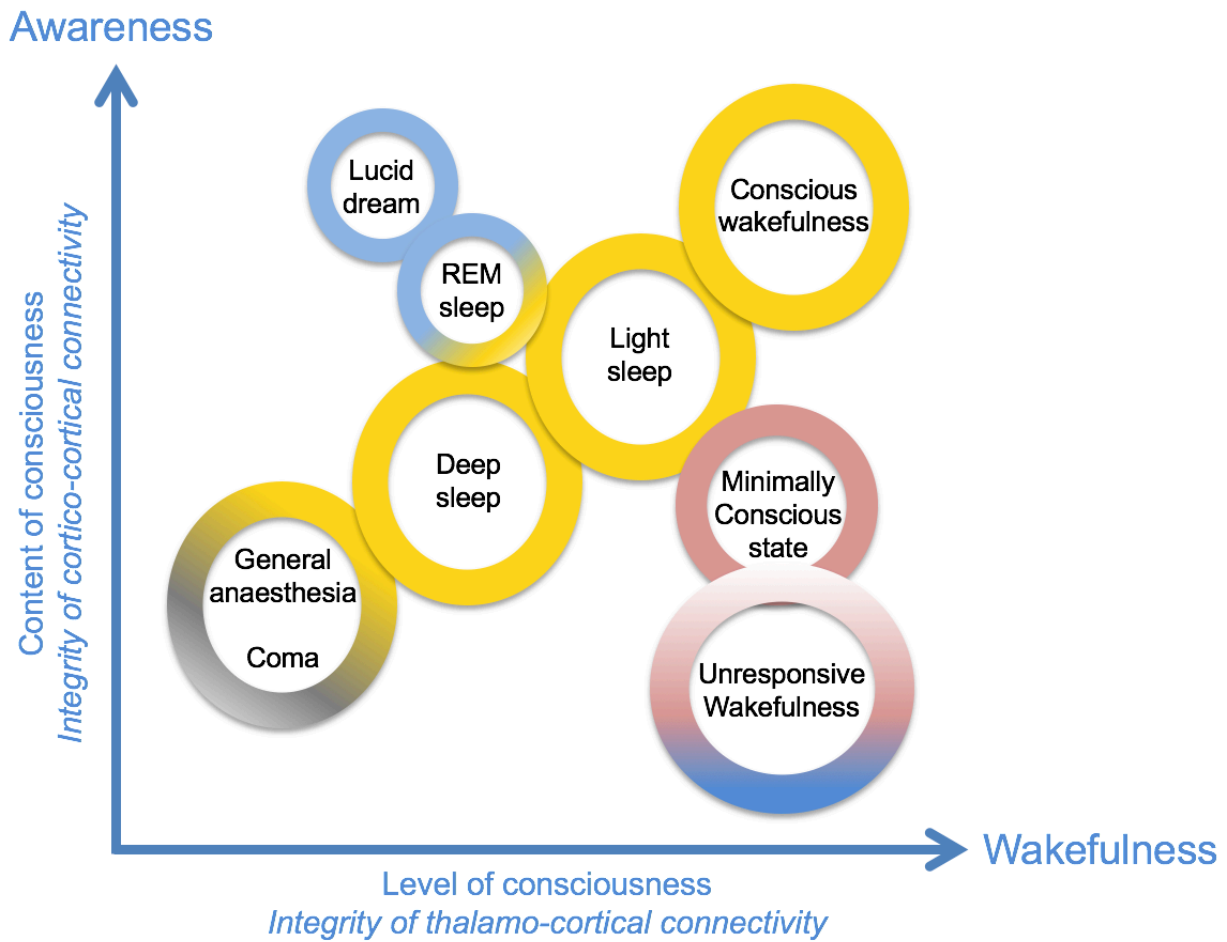

**Figure 1**. **Wakefulness and awareness are two essential dimensions of consciousness.** In this diagram, several qualitatively different states of consciousness have been positioned on the 2D-matrix as a function of the associated axes "content of consciousness" (awareness) and "level of consciousness" (wakefulness). Adapted from (Laureys, 2005).

## 2. Functional networks, a flexible architecture for conscious processes

The main disadvantage of static network architectures is their limitation in terms of amount and variety (complexity) of information processing that can take place. The brain takes advantage of different mechanisms that overcome this limitation, by enabling a dynamic, transient reconfiguration of brain networks increasing the repertoire of possible responses to inputs (i.e., complexity of input-output functions) (Sporns, 2013). Such transient networks involving only a few brain regions, coordinated to achieve a specific function limited in duration, form what is termed functional connectivity. There is a growing interest regarding the functional networks associated with specific cognitive tasks (Hassan et al., 2015) and novel frameworks have recently emerged (Avena-Koenigsberger et al., 2017) to explain how brain-scale anatomical connectivity relates to functional connectivity. Functional networks organize through the network "means of communication", also termed communication dynamics, that governs information routing through specific networks, instead of propagating information through the entire brain network (Avena-Koenigsberger et al., 2017). If that was the case, then information generated locally would induce distant activity in all connectome-related regions, resulting in a massively synchronized response with low informational content and complexity.

More specifically, one fundamental question is: what are the mechanisms regulating communication dynamics and enabling functional networks to emerge in brain-scale networks? This question is central to understand how the brain optimizes its information processing capabilities, which are tightly linked with consciousness. We propose that the fundamental mechanisms underlying communication dynamics are actually cellular-scale mechanisms that 1) prevent brain-scale neuronal synchronization following a stimulus, and 2) enable the transient coupling of specific distant brain regions. There has been a considerable amount of interest for large-scale brain activity patterns linked with consciousness, since those can be measured through various neuroimaging modalities (e.g., electroencephalography, EEG; functional magnetic resonance imaging, fMRI). However, mechanisms at the cellular scale have remained more challenging to address in humans for obvious reasons of invasiveness associated with the required recording techniques in humans. In this section, we review evidence for such mechanisms that could bridge the micro-circuit and brain-scale levels.

In terms of large-scale neuroanatomical pathways enabling consciousness, long-range glutamatergic projections between pyramidal neurons through white matter fibers have likely a critical role (Dehaene and Changeux, 2011) since they enable fast (due to myelin) communication between distant regions. At the brain-scale level, these white matter fibers are likely critical to enable conscious access, which involves the transient stabilization of neuronal activity encoding a specific information pattern, in a network of high-level brain regions interconnected by long-range connections, with the prefrontal cortex (PFC) acting as a key node (Dehaene et al., 2006; Berkovitch et al., 2017). On conscious trials, distributed gamma-band activity reflects a stabilization of local information broadcasted to other areas. Global broadcasting is thought to make the information accessible to introspection and reportable to other brain regions (Lamme, 2010). During conscious access to a specific information, other surrounding global workspace neurons would be inhibited and unavailable for the processing of other stimuli, therefore remaining preconscious (not reaching consciousness).

At the local scale, a micro-circuit has also been identified as being involved in the communication between distant brain regions: the projection from pyramidal neurons in a brain region to VIP-positive (Vasoactive Intestinal Peptide) neurons in another region. By activating VIP-positive neurons in a distant region, this induces an inhibition of somatostatin-positive (SST) neurons, which target pyramidal cells dendrites, resulting in a disynaptic disinhibition (Karnani et al., 2016). Through this disynaptic disinhibition, gamma activity generation can occur through PV-PV mutual inhibition, and binding between the two involved regions can possibly take place (Munoz et al., 2017), temporarily enabling information transfer and processing. These cellular-scale mechanisms are summarized in Figure 2.

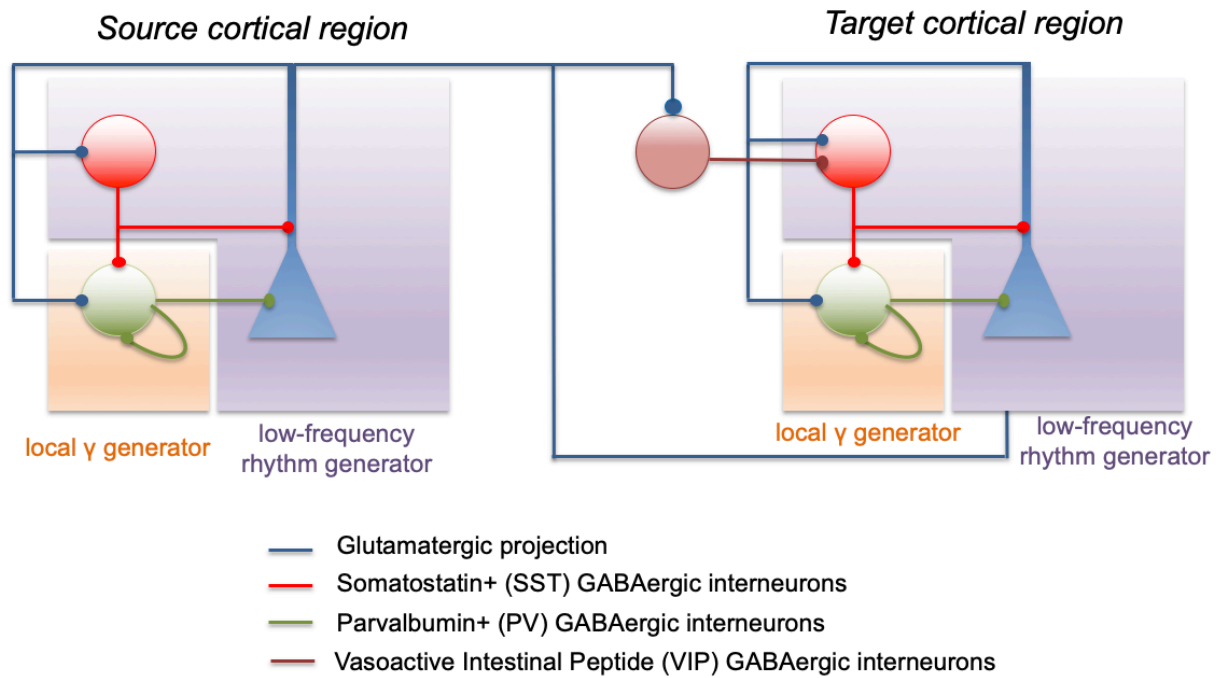

**Figure 2**. **Basic (schematic) circuits involved in the generation of local and distant oscillations.** In a local cortical network (cortical column), gap-junctional, mutual inhibition of soma-projecting, "fast" GABAergic interneurons are one of the basic mechanisms of generation for local gamma activity, along with the PING (Pyramid-InterNeuron Gamma). Conversely, the feedback loop between dendrite-projecting, "slow" GABAergic interneurons and pyramidal cells can generate low-frequency activity. Importantly, distant communication through the disynaptic pathway enabling transient generation of gamma oscillations in distant populations. Pyramidal cells in the source population (left circuit) project on the pyramidal cells of the distant population (right), but also on VIP interneurons that project on dendrite-projecting SST neurons. Transient activation of VIP neurons from the source population transiently inhibits SST neurons in the target population, enabling the generation of gamma oscillations through the PV-PV and PYR-PV circuit. Once the input on distant VIP neurons decreases, SST neurons resume their inhibitory input, which can terminate gamma oscillations generation.

Considering these cellular-scale mechanisms, one crucial question is to understand how they are involved into the transient emergence and maintenance of functional networks such as those supporting consciousness. It is established that the state of consciousness is critically dependent on brain functional cortico-cortical connectivity (Marino *et al.*, 2016; Naro *et al.*, 2018). As a reminder, *functional* connectivity refers to statistically significant couplings between temporal courses of neuronal activity within different regions while *anatomical* connectivity denotes the physical connections between brain regions. Functional networks can therefore reflect indirect connections between brain regions, and are transient depending on which tasks are performed, or which stimuli are perceived. Even in the absence of any specific task or stimuli, it has been shown that resting-state networks (e.g., default mode network -DMN-) are also transient (Kabbara *et al.*, 2017). Experimental evidence supports the idea that functional connectivity can shed light on the networks involved in various conscious states (Jin and Chung, 2012). For example, during general anaesthesia-induced loss of consciousness, there is a breakdown in cortico-cortical functional connectivity (Ferrarelli *et al.*, 2010; Hudetz, 2012; Gomez *et al.*, 2013), severely impairing the capacity of cortical networks to integrate information and to make it available at a large scale, as required for conscious perception in IIT or GWT. Similarly, in the transition from wakefulness to slow-wave sleep, the firing rate in the cortex remains relatively unchanged during the depolarizing phases of the slow sleep oscillation (Steriade *et al.*, 2001), while effective brain connectivity is dramatically altered (Tononi and Sporns, 2003; Esser *et al.*, 2009). Upon falling into NREM sleep, cortical activations also become more local and stereotypical, impairing effective cortical connectivity (Massimini *et al.*, 2010), as shown using TMS-evoked EEG responses which remain very close to the stimulation site; while these responses involve a network of distant brain regions undergoing complex dynamical patterns of successive activation during wakefulness (Guillery and Sherman, 2002; Casali *et al.*, 2013; Casarotto *et al.*, 2016). These results also emphasize the crucial role of the thalamo-cortical pathway in cortico-cortical functional connectivity. It is worth noting that the essential role of the thalamocortical loop as well as so-called "reentrant interactions" were previously considered as key in the DCH (Tononi and Edelman, 1998).

Consistently with these results obtained during sleep, this breakdown of cortico-cortical connectivity has also been observed during general anaesthesia and in DOC patients, and even explored through computational modeling (Esser *et al.*, 2009; Casali *et al.*, 2013). In the GWT framework, this explains why consciousness is impaired in such states: large-scale communication between distant brain areas is impaired due to thalamo-cortical modulation, preventing ignition from occurring. In brain-damaged DOC patients, large-scale cortico-cortical communication can be impaired through the partial destruction of long-range fibers, physically impeding long-range brain synchrony. In terms of effects at the cellular scale, destruction of long-range fibers could prevent the synchronization of distant VIP interneurons, which is critical to induce disynaptic disinhibition and associated gamma activity required for CTC.

Pathological alterations of functional connectivity have been investigated using a variety of modalities: (1) functional connectivity during ''resting state'' using fMRI or EEG; (2) pulsed stimulation using transcranial magnetic stimulation (TMS) during EEG recording; and (3) other perturbation-based approaches investigating brain responses to sensory stimuli (Boly et al., 2017). The advantage of functional connectivity is that it can be employed to improve the evaluation and classification of disorders of consciousness (Sanders et al., 2012; Holler et al., 2014; Rossi Sebastiano et al., 2015; Naro et al., 2018). For example, in mild cognitive impairment (MCI) patients, it has been shown that impaired consciousness is associated with altered effective connectivity (Varotto et al., 2014; Crone et al., 2015). Failure of large-scale connectivity, along with a hypersynchrony of local short-range delta and alpha activity were detected within the DMN and were correlated with the level of awareness in patients with DOC (Vanhaudenhuyse et al., 2010; Fingelkurts et al., 2013; Maki-Marttunen et al., 2013; Varotto et al., 2014; Kabbara et al., 2017; Naro et al., 2018). Furthermore, the functional connectivity pattern of several brain regions, such as the posterior cingulate cortex and precuneus, may even predict UWS patients' state improvement of consciousness with an accuracy superior to 80% (Wu et al., 2015).

Beside "passive" investigation of resting-state functional connectivity, the use of TMS-evoked EEG responses enables the active "probing" of functional connectivity. For example, a drastic breakdown of functional connectivity has been identified in UWS patients using a specific TMS protocol triggering, in these patients, a stereotyped, local EEG response similar as in unconscious sleeping or anaesthetized subjects (Rosanova et al., 2012). Restoring cortical large-scale effective connectivity with transcranial brain stimulation, such as tACS, in DOCs could therefore be a useful approach to facilitate partial recovery by enhancing oscillations and plasticity. One clinical result supporting this idea is the recent demonstration that DLPFC (dorsolateral prefrontal cortex)-tACS was able to transiently restore the connectivity breakdown in DOC individuals (Naro et al., 2016).

One fundamental microscopic-scale mechanism involved in information routing in the brain, and contributing to form functional networks within the anatomical network, is called Gating By Inhibition (Jensen and Mazaheri, 2010). GBI involves inhibitory processes resulting in the selective activation of sub-networks and inactivation of other sub-networks. By preventing brain-scale activation in response to a stimulus, and restricting the number of brain regions engaged in performing tasks, GBI also prevents states of low complexity (e.g., all brain regions displaying the exact same activity) and therefore inefficient information processing. Thus, GBI processes suggest that the role of inhibition is more complex than preventing excessive activation of brain networks, contributing instead to shaping anatomical brain networks into functional networks (Avena-Koenigsberger et al., 2017). Possible alterations of GBI were reported in studies showing that EEG alpha power is lower in UWS than in MCS patients (Lehembre et al., 2012; Stefan et al., 2018), hinting that the neurobiological mechanisms underlying alpha oscillations generation and associated GBI are profoundly altered in unresponsive patients. Moreover, alpha activity was highly synchronized and clustered in central and posterior cortical regions in UWS patients (Lehembre et al., 2012; Stefan et al., 2018), suggesting a possible failure of GBI in the most severe disorders of consciousness.

## 3. Information processing in large-scale functional networks through nested oscillations

One of the most established processes by which distant brain regions engage together in an activity pattern associated with the performance of a given task is Communication Through Coherence (CTC) (Fries, 2005, 2009, 2015b; Deco and Kringelbach, 2016; Bonnefond et al., 2017). CTC involves indeed phase-coupled gamma activity between distant brain regions to enable information processing. CTC has been suggested to be the substrate of "binding", i.e. the merging of different features of a stimulus into a single, unified conscious perception (Singer, 2001). More precisely, the excitability fluctuation in a group of neurons provides a specific signature characterized by a specific frequency band and pattern of discharge (Womelsdorf et al., 2014), propagating through a large-scale network consisting of anatomically interconnected brain areas and subsequently triggering activity in connected regions. Information processing in the brain is strongly linked with phase-locked, coordinated-in-time fluctuations of excitability (Fries, 2005; Fries, 2015a) in networks of distributed neuronal populations. The resulting oscillations generate a specific neuronal code, and coherence enables the association of information and communication. Furthermore, CTC involves gamma activity, generated mainly by GABAergic interneurons (PV-positive basket cells). Taken together, inhibitory processes appear key for information routing and processing in brain-scale networks involved in consciousness: GBI shapes brain networks spatially (which brain regions are involved, and which ones are inhibited), while CTC controls them temporally (information flow). However, this raises an intriguing question: if gamma rhythms are generated locally by interneuronal - GABAergic- networks, how can distant brain regions, connected through glutamatergic long-range fibers, communicate efficiently and achieve CTC? One possibility is that the co-occurrence of low- and high-frequency neuronal oscillations could provide distant co-activation (low-frequency, glutamatergic origin) that would then enable binding (high-frequency, GABAergic origin). This would involve the formation of a functional network of several brain regions through the low-frequency rhythm, prior to information transfer and processing through CTC (involving gamma activity).

As a support for this possibility, conscious perception is indeed characterized by an increase in distributed gamma-band activity (Melloni *et al.*, 2007; Wyart and Tallon-Baudry, 2009). Interestingly, these fast oscillations are modulated by slow oscillations (Osipova *et al.*, 2008; Jensen *et al.*, 2014). It has recently been proposed that phase synchronization of low-frequency oscillations, playing the role of a temporal reference frame for information, carrying high-frequency activity, is a general mechanism of brain communication (Bonnefond *et al.*, 2017). These nested oscillations might be a key mechanism, not only for cortico-cortical communication and processing, but also between sub-cortical structures. Emotional memory, involving both cortical and subcortical structures, indeed engages large network synchronization through nested theta-gamma oscillations (Bocchio *et al.*, 2017). During *in vivo* experiments performed in rodents, a perceived threat (a stimulus announcing a footshock) enhances theta power and coherence in the amygdala, prefrontal cortex, and hippocampus (Lesting *et al.*, 2011; Likhtik *et al.*, 2014), while fast gamma bursts are phase-locked to theta oscillations (Stujenske *et al.*, 2014). Overall, these results support the idea that nested oscillations at theta and gamma frequencies are a plausible substrate for information channel opening/routing (theta) and processing/transfer (gamma) within the brain.

Attention is another key element for conscious processing and it is involved in the synchronization of distant brain regions (Steinmetz *et al.*, 2000; Niebur *et al.*, 2002). The main underlying brain rhythms involved in attentional processes are alpha and gamma oscillations: brain regions synchronize gamma oscillations (Womelsdorf *et al.*, 2014), and are modulated by slow alpha oscillations. Slow oscillations enable inhibition of irrelevant networks, influence local signal processing, widespread information exchange, and perception (Sadaghiani and Kleinschmidt, 2016). Information flow is established by neuronal synchronization at the lower-frequency bands, namely in the theta (4–7 Hz), alpha (8–13 Hz), and beta (14–25 Hz) bands (Bonnefond *et al.*, 2017). One possible reason is that low-frequency activity induces a transient change in excitability in target brain structures, which provides an optimal window for binding neuronal signals from different regions through high-frequency activity (i.e., gamma) (Canolty *et al.*, 2006). This provides further support to the idea that low-frequency neural oscillations are mainly involved in establishing transient long-range communication through glutamatergic projections, while high-frequency neural oscillations are rather involved in information processing/transfer. It is therefore possible to relate the notion of "integration" with this long-range, glutamatergic co-activation, enabling brain-scale communication between brain regions; while "differentiation/segregation" would rather depend on locally generated gamma activity (and in part on low-frequency activation level, which would result in massively synchronized activity, and reduced differentiation and complexity). An overview of the aforementioned mechanisms is proposed in Figure 3.

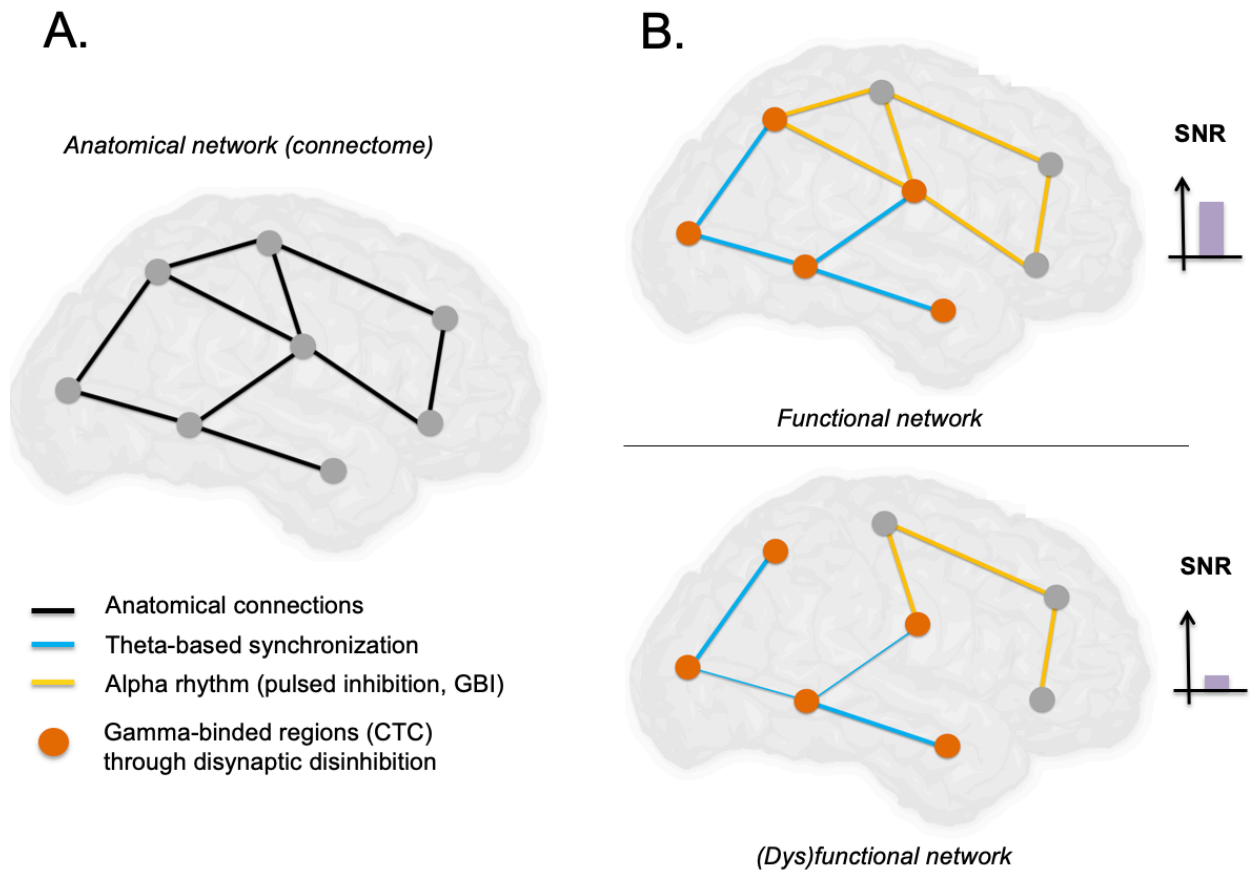

**Figure 3**. **Schematic overview of transient, selective binding among cortical networks through cellular-scale mechanisms. A.** Schematic diagram of an anatomical network with main projections between regions at the brain scale. **B.** *Upper panel.* Selective binding in subset of cortical regions occurs through the generation of gamma

oscillations (mostly through micro-circuits involving basket cells), while the distant disinhibition of specific brain regions occurs through disynaptic disinhibition (activation of distant VIP neurons inhibits SST neurons, which in turn decrease their inhibitory projection on pyramidal neurons). This contributes to shape the anatomical network into a functional network. The alpha rhythm acts as a pulsed inhibition to inhibit "irrelevant" networks, increasing further the signal-to-noise (SNR) ratio. *Lower panel.* Decreased integration (e.g., following brain damage) leads to impaired synchronization of distant brain regions (reflected by decreased low-frequency rhythm on the illustrative oscillation), and thereby a decrease of binding, which combined to a decrease in GBI efficiency strongly decreases the overall SNR, leading to dysfunction of the network in terms of integration and binding required for consciousness. Let us note that, in addition to decreased amplitude of the low-frequency rhythm, the phase relationship of the nested gamma oscillation could be perturbed (i.e., more random) as compared to the physiological case. Such potential relationship of nested theta/gamma oscillations remains to be explored in DOC.

Emerging evidence shows that the local versus global information processing balance can be impaired in neurological disorders. Typically, a recent study investigating functional networks in Alzheimer's disease patients identified a decrease in brain integration as quantified by the participation coefficient (reflects communication between distant brain modules), while segregation as quantified by the clustering coefficient (reflects local communication between neighbour brain regions) was increased (Kabbara et al., 2018). This is consistent with neurodegenerative processes, which likely impact the "locking" of specific brain regions or the inhibition of irrelevant networks, thereby severely impairing large-scale integration of information. In the context of DOC, recent clinical evidence (Chennu *et al.*, 2017; Rizkallah *et al.*, 2019) supports this view. In the study by Chennu et al., scalp-level networks were assessed from DOC patients and pointed at decreased integration within the alpha band. More specifically, the fronto-parietal network in the alpha band was discriminant between MCS and UWS patients. In a recent study, Rizkallah et al. (Rizkallah *et al.*, 2019) quantified the level of local versus global information processing in frequency-dependent functional networks (source level) in DOC patients and controls. Integration in theta band functional networks decreased with consciousness level, and two anatomical regions were systematically involved between controls and any patient group: a portion of the left orbitofrontal cortex and the left precuneus. One possibility is that physical damage to long-range white matter fibers impairs large-scale integration and the local/global information processing balance. One possible approach to study such anatomical damage to white matter fibers is diffusion tensor imaging (DTI), as performed in DOC patients suffering severe brain injury (Fernandez-Espejo et al., 2011; Galanaud et al., 2012; Luyt et al., 2012), which highlighted widespread disruptions of white matter. Lower fractional anisotropy was indeed found in the subcortico-cortical and cortico-cortical fiber tracts of DOC patients as compared to controls (Lant et al., 2016; Weng et al., 2017), suggesting that major consciousness deficits in DOC patients may be related to altered WM connections between the basal ganglia, thalamus, and frontal cortex. This is also in line with the effect of lesion of myelinated fiber tracts, which can result in a failure of communication between distant brain regions (Adams et al., 2000). Therefore, it seems reasonable that white matter lesions can alter, modify or prevent both CTC and GBI between large-scale networks. Furthermore, we speculate that, should the specific phase-locking of gamma oscillations onto theta oscillations be perturbed, then clinical manifestations associated with DOC might appear (loss of integration and decrease in the consciousness level).

## 4. Possible clinical implications

In this review, we have attempted to reconcile the neuroanatomical and neurophysiological knowledge at the level of micro- and macro-scopic networks, regarding the processes that underlie the emergence and maintenance of consciousness, and its alterations in DOC patients.

The multiplexing of neuronal rhythms through nested oscillations appears as a plausible mechanism of co-activation in a network of specific distant brain regions (integration), which is a pre-requisite for a conscious perception. Furthermore, a key mechanism seems to be the subtle balance between low-frequency activity (associated with "global" processing) and high-frequency activity (associated with a more "local" processing), which could enable the neuronal dynamics underlying optimal information routing and processing. Excessive low-frequency activity (e.g., delta activity) results in massively, synchronized activity resulting in a loss of complexity in terms of information processing, paralleled in such cases with a loss of consciousness (e.g., sleep, seizures). Similarly, a lack of fronto-parietal functional coupling (attentional network) has been recently observed in a recent study, as quantified using high-resolution EEG in DOC patients (Chennu et al., 2017), suggesting that a sufficient level of fronto-parietal coupling is required to achieve sufficient neuronal integration and ignition for conscious perception. More generally, brain dynamics in DOC patients is typically characterized by a loss of integration at a large-scale (Chennu *et al.*, 2017; Rizkallah *et al.*, 2019), preventing efficient large-scale coordination of distant brain regions to achieve conscious perception. This suggests that the low-frequency rhythm required for long-range cortical communication is decreased, preventing binding in the gamma range and therefore further processing information and ignition for consciousness access. That being said, what are the possible implications of this slow/fast activity balance as a candidate mechanism for the complex processing associated with consciousness?

An interesting perspective, which would also be a form of validation for this mechanism, is the use of neuromodulation techniques in DOC patients to increase their level of consciousness. The objective of such neuromodulation techniques could be to "re-balance" local versus global processing, for example through the use of transcranial direct or alternating stimulation (tDCS/tACS), applied to both a frontal and a parietal site simultaneously, in order to increase low-frequency synchronization in the theta range. It is the current view that tACS can modulate endogenous brain rhythms using relatively low-levels electric fields (<1 V/m, as discussed in (Modolo *et al.*, 2018), which would involve to use a stimulation frequency in the theta range to increase residual oscillations in this frequency range (i.e., assuming that residual anatomical connections are still present). Dual-site, fronto-parietal tACS in the theta range could then provide a non-invasive possibility to increase the level of consciousness in DOC patients, pending that some residual anatomical connectivity remains in the case of brain-damaged patients. Interestingly, a recent study (Violante et al., 2017) used dual-site tACS in the theta range in healthy volunteers reported improved working memory, a function also dependent on fronto-parietal networks. Another study, using tACS targeting the fronto-temporal network, reported an increase in working memory performance in seniors to comparable levels than young participants (Reinhart and Nguyen, 2019). These recent results, obtained in humans, provide compelling evidence that re-balancing information processing through neuromodulation protocols could contribute to increase the level of consciousness in some DOC patients (e.g., where damage to long-range white matter fibers is not too severe).

**Discussion and concluding remarks**

The two dimensions of consciousness, awareness and wakefulness, depend on anatomically distinct pathways: cortico-cortical "horizontal" connectivity and thalamo-cortical "vertical" connectivity. Excessive "up-and-down-like" thalamo-cortical activity impairs cortico-cortical connectivity by due to excessive lateral inhibition, thereby preventing CTC of distant brain regions, resulting in drastically altered functional connectivity, in line with the loss of consciousness in deep sleep or DOC. This control of cortico-cortical communication by thalamo-cortical activity is fundamental in understanding how attentional processes can emerge by transiently recruiting efficiently, through nested low-(theta) and high-(gamma) frequency rhythms, distant brain regions. The evidence reviewed highlights how the balance of nested brain rhythms with fundamentally different functions can transform an anatomical network into a transient, successive activation of different sub-networks, i.e. a functional network. This versatility of reconfiguration of the structural connectome results in an immense and complex dynamical repertoire of functional networks with specific rhythms and cross-frequency couplings, probably a key infrastructure enabling consciousness. Among those rhythms, three appear especially involved in conscious processes: while the function of the theta rhythm appears linked with "opening" transient channels of communications through distant regions, the alpha rhythm seems to play the role of pulsed inhibition to increase further the SNR. There is also solid and converging evidence that gamma oscillations are an excellent candidate for information processing and transfer.

Importantly, mechanisms identified at the micro-circuit scale between specific types of interneurons, such as projections from pyramidal cells to distant VIP cells, are critical to provide a more mechanistic framework for theories of consciousness, notably GWT. Such mechanisms indeed clarify the conditions under which ignition can occur, while providing links with other concepts that are not necessarily unified (e.g., CTC, GBI) to enable access to consciousness. In addition to the recruitment of selective brain regions to have access to the global workspace, it also appears important to take into account that active inhibitory processes co-occur to improve the signal-to-noise ratio (e.g., GBI). An overview of those concepts, along with the contribution of the reviewed mechanisms to the increase in neural activity complexity associated with consciousness, is provided in Figure 4.

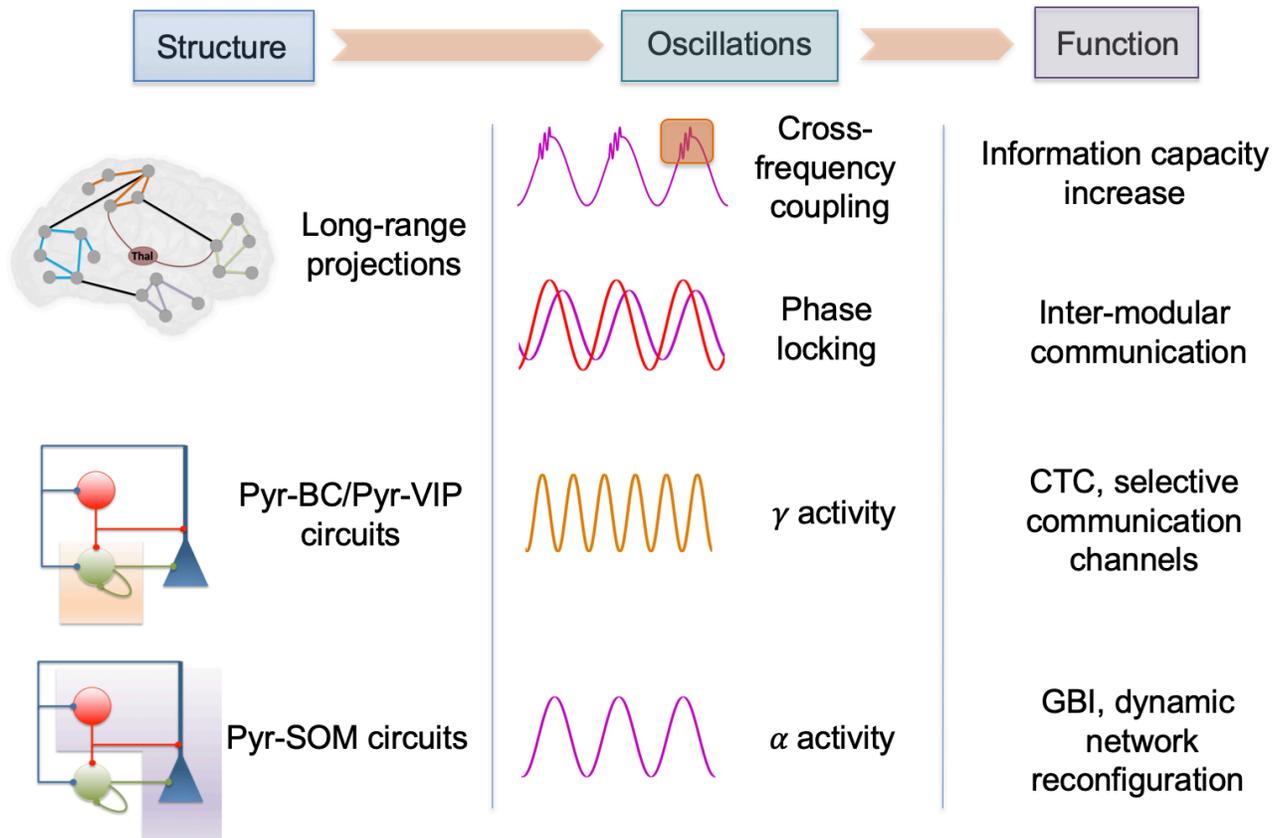

**Figure 4.** Synthesis of the network-level mechanisms underlying the complexity associated with conscious processes. The structural characteristics of brain circuits at different scales are mentioned with some key oscillatory rhythms and associated functions.

Taken together, the mechanisms presented in this review suggest that the importance of the thalamo-cortical pathway, emphasized in the DCH theory, cannot be neglected within the context of the GW theory: the thalamo-cortical pathway plays actually the role of a "switch", enabling or not efficient integration and communication within cortico-cortical networks through feed-forward inhibition. Therefore, efficient modulation of the thalamo-cortical pathway is necessary, but not sufficient, to enable ignition within cortical networks and availability of information at a large scale. For these reasons, it seems that DCH and GWT are both accurate each from their perspective, and could be unified to obtain a more integrated vision, through a new framework accounting for both aspects (ignition, availability of information, and control/routing of information by thalamo-cortical pathways). In such framework, two balances are critical: *the first one* is between vertical (thalamo-cortical) and horizontal (cortico-cortical) connectivity, which controls *the second one* between local and distant information processing within cortico-cortical networks. Consequently, we propose that DCH and GWT could be reconciled through this balance between horizontal and vertical connectivity, and account for a wider range of phenomena related to consciousness and its deregulations.

**Future directions**

- An important step forward would be to investigate further the cross-frequency coupling between the low-frequency theta rhythm and high-frequency gamma rhythm in healthy controls as compared to DOC patients during resting state. The identification of such changes could have implications in terms of diagnostic evaluation, but also regarding novel neuromodulation protocols that might aim, at least in part, to regulate abnormal cross-frequency couplings.

- Another promising application would be to translate the circuitry presented in this review into a tractable computational model consisting in a network of brain regions, possibly using the neural mass model approach. Evaluating *in silico* how the microcircuits are involved into the generation of nested theta-gamma oscillations, and how TMS-evoked EEG responses at the brain scale are impacted by synchronized thalamocortical activity, would provide key mechanistic understanding. Candidate tDCS/tACS protocols could also be tested and evaluated *in silico*.

- Characterizing further the dynamics of functional networks in DOC patients using EEG, for example by studying the nature and dynamics of modular states over time (e.g., dwell time) as a function of the level of consciousness (wake, sleep, DOC such as UWS). Extracting such dynamical information about functional brain network could have diagnostic implications, notably to distinguish between MCS and UWS patients.

## Competing financial interests statement

The authors have no competing financial interests to declare for this work.

## Funding

This study is funded by the Future Emerging Technologies (H2020-FETOPEN-2014-2015-RIA under agreement No. 686764) as part of the European Union's Horizon 2020 research and training program 2014–2018.